\documentclass[onecolumn, 14pt]{IEEEtran}
\usepackage[english]{babel}
\usepackage{graphicx}
\usepackage{booktabs}
\usepackage{rotating}
\usepackage{listings}
\usepackage[cmex10]{amsmath}
\usepackage{amsfonts}
\usepackage{amssymb}
\usepackage{morefloats}
\usepackage{cite}
\usepackage{bm}
\usepackage{setspace}
\begin{document}

\title{Automated Segmentation of Brain Gray Matter Nuclei on Quantitative Susceptibility Mapping Using Deep Convolutional Neural Network}

\author{Chao Chai, Pengchong Qiao, Bin Zhao, Huiying Wang, Guohua Liu, Hong Wu, E Mark Haacke, Wen Shen, Chen Cao, Xinchen Ye, \IEEEmembership{Member, IEEE}, Zhiyang Liu, \IEEEmembership{Member, IEEE}, Shuang Xia
\thanks{This work was supported by the Natural Science Foundation of China (NSFC) under Grant nos. 81901728, 61871239, 81871342 and 81873888, and National Key Technologies Research and Development Program of China  under Grant no. 2019YFC0120901.}
\thanks{C. Chai, H. Wang, W. Shen and S. Xia are with Department of Medical Imaging, Tianjin First Central Hospital, Tianjin, 300192, China, and Tianjin Medical Imaging Institute, Tianjin 300192, China.}
\thanks{P. Qiao, B. Zhao, G. Liu, H. Wu and Z. Liu are with Tianjin Key Laboratory of Optoelectronic Sensor and Sensing Network Technology, College of Electronic Information and Optical Engineering, Nankai University, Tianjin 300350, China.}
\thanks{E. M. Haacke is with Department of Radiology, Wayne State University, Detroit, Michigan 48202, USA.}
\thanks{C. Cao is with Department of Medical Imaging, Tianjin Huanhu Hospital, Tianjin 300350, China.}
\thanks{X. Ye is with DUT-RU International School of Information Science and Engineering, Dalian University of Technology, Dalian 116620, China.}
\thanks{Corresponding authors: Z. Liu (email: liuzhiyang@nankai.edu.cn) and S. Xia (email: xiashuang77@163.com)}
}
\date{}
\maketitle
\begin{abstract}
    Abnormal iron accumulation in the brain subcortical nuclei has been reported to be correlated to various neurodegenerative diseases, which can be measured through the magnetic susceptibility from the quantitative susceptibility mapping (QSM). To quantitively measure the magnetic susceptibility, the nuclei should be accurately segmented, which is a tedious task for clinicians. In this paper, we proposed a double-branch residual-structured U-Net (DB-ResUNet) based on 3D convolutional neural network (CNN) to automatically segment such brain gray matter nuclei. To better tradeoff between segmentation accuracy and the memory efficiency, the proposed DB-ResUNet fed image patches with high resolution and the patches with low resolution but larger field of view into the local and global branches, respectively. Experimental results revealed that by jointly using QSM and T$_\text{1}$ weighted imaging (T$_\text{1}$WI) as inputs, the proposed method was able to achieve better segmentation accuracy over its single-branch counterpart, as well as the conventional atlas-based method and the classical 3D-UNet structure. The susceptibility values and the volumes were also measured, which indicated that the measurements from the proposed DB-ResUNet are able to present high correlation with values from the manually annotated regions of interest.
    \end{abstract}

\section{Introduction}

Tissue magnetic susceptibility is a physical magnetic resonance imaging (MRI) parameter that indicates how the local magnetic field changes in tissues when an external magnetic field is applied. Tissue magnetic susceptibility is able to reflect the unique information about tissue composition including the iron and myelin\cite{li_association_2015, Chai2019}. It has been reported that focal abnormal iron accumulation in brain has been observed in various neurodegenerative diseases, such as Alzheimer's Disease\cite{Gong2019}, Parkinson's Disease\cite{Xuan2017} and Huntington's Disease\cite{Chen2019}. The abnormal brain iron deposition in these diseases is prone to occur in the subcortical nuclei including the caudate nucleus (CN), globus pallidus (GP), putamen (PUT), thalamus (THA), substantia nigra (SN), red nucleus (RN), and dentate nucleus (DN). These subcortical nuclei were involved in executive functions and motor control, such as behavioral control, emotion, and motor learning\cite{li_association_2015, florio_basal_2018}. Owing to its ability in quantitively measuring the magnetic suspectibility, the quantitative susceptibility mapping (QSM)\cite{liu_susceptibility-weighted_2015} has shown to be an important non-invasion measurement method in monitoring neurodegenerative diseases, and presented great potential in test new therapies or drugs.

Currently, to quantitatively measure the magnetic susceptibility of subcortical nuclei from the QSM, the regions of interest (ROIs) were mostly obtained from manual delineation, which was a tedious and labor-intensive work, and the delineation was also heavily dependent on the evaluators' experience. Therefore, to improve both the efficiency and the accuracy, it is urgent to develop a segmentation method in an automatic way. Conventionally in medical imaging, the automatic nuclei segmentation methods were mostly atlas based\cite{igual_fully-automatic_2011, igual_automatic_2012,xia_automatic_2007, su_thalamus_2019,li_multi-atlas_2019}, where an atlas of the targeted ROIs were annotated on a standard brain, typically in the Montreal Neurological Institute (MNI) space. To segment the ROIs, the target images were first registered with the standard brains by using T$_\text{1}$ weighted imaging (T$_\text{1}$WI) or T$_\text{2}$weighted imaging (T$_\text{1}$WI) with thin slice thicknesses to obtain a transform between the subject-in-question's brain and the standard brain. After annotating the ROIs according to the templates, an inverse transform was applied to the ROI segmentation maps to obtain the segmentation results. Due to the dissimilarities between the subject-in-question's and the standard brain, one of the key challenges is how to obtain an perfect registration between the subject-in-question's and the standard brains, so as to assure the accuracy of the ROI templates. Another challenge comes from the fact that the the appearances of the gray matter nuclei are highly correlated to the anatomical structures, which may vary due to aging, tumor or other types of diseases. To tackle these challenges, probabilistic models were usually adopted to improve the segmentation performance \cite{li_multi-atlas_2019, su_thalamus_2019}. By using several subjects with both sexes and various ages, the segmentation method becomes more robust against the anatomical dissimilarities, which further increase the segmentation accuracy.

\begin{figure}
    \centering
    \includegraphics[width=0.7\linewidth]{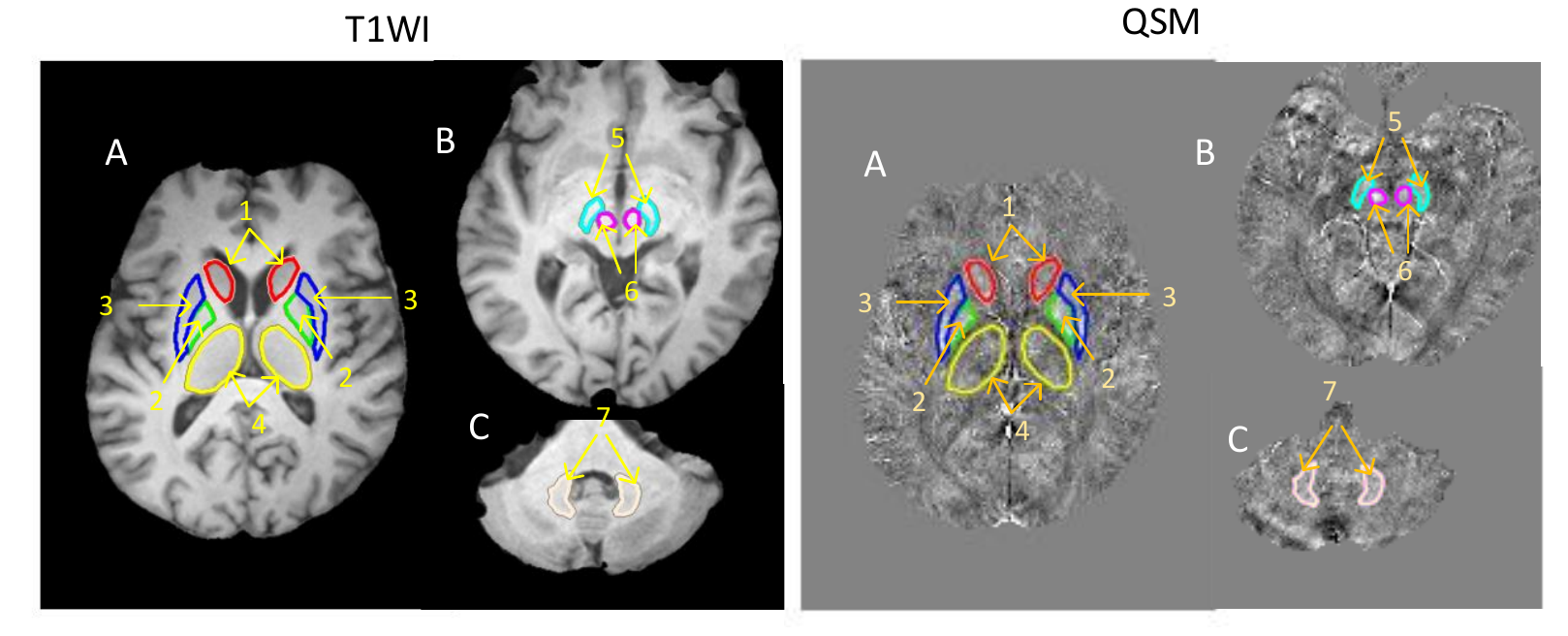}
    \caption{The regions of interest of different gray matter nuclei outlined on both the QSM and T$_\text{1}$WI. (A) The level of basal ganglia. (B) The level of mid-brain. (C) The level of cerebellum. 1, CN; 2, GP; 3, PUT, 4, THA; 5, SN; 6, RN; and 7, DN.}
    \label{fig:nuclei}
\end{figure}

\begin{figure}
    \centering
    \includegraphics[width=\linewidth]{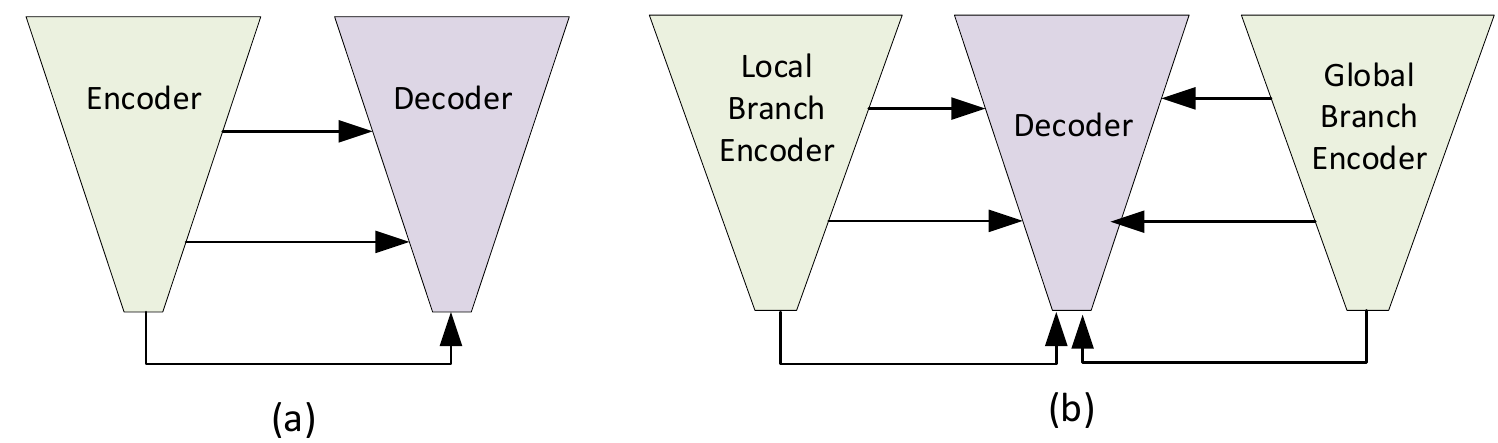}
    \caption{Overall pipeline of deep-learning-based segmentation methods. (a) U-Net like structure. (b) Our proposed local and global fusing network structure.}
    \label{fig:basic_structure}
\end{figure}

Another possible way to solve the imperfect registration and the dissimilarities between the subject-in-question and the MNI space is to avoid registering with the standard brain, which requires that the algorithm \textit{understands} the tissue-to-segment. Deep learning has recently been adopted in computer vision and medical image processing thanks to its strong ability in extracting features from big data. By using a large amount of labeled data samples, the deep artificial neural network is able to learn effective representation of features that improves the performance on specific tasks. The rapid development in the parallel computing capability of graphic processing unit (GPU) further accelerates the training a complicated neural network structure on a massive number of data samples. In image classification task, for instance, the accuracy of the convolution neural network (CNN) has surplus the average performance of human. The strong power of CNN in learning representation from images should contribute most to the convolution layers\cite{lecun_deep_2015}. Despite that deeper network has been shown to be more expressive than its shallower counterpart, simply stacking many convolution layers would impose significant difficulties in training due to the so-called gradient-vanishing problem. To build deeper and trainable network, by introducing residual structures to the CNN, ResNet is able to build very deep networks, and achieved remarkable performance in image recognition tasks\cite{he_deep_2015}. With residual structures, the networks have no spurious local optima, making it easier to converge \cite{hardt_identity_2018}.

Basically, the well-performed segmentation networks \cite{chen_deeplab_2017, chen_encoder-decoder_2018} were mostly designed for 2D natural images. In medical image segmentation, the U-Net structure \cite{ronneberger_u-net_2015} and its deformations have achieved tremendous performance in 2D images segmentation tasks, such as cell nuclei or cell boundaries segmentation from histopathological images \cite{isensee_nnu-net_2018, dolz_ivd-net_2019}. When processing 3D images such as CT and MR images, despite that the methods developed for 2D images can be directly applied by treating the 2D slices as independent images, the segmentation performance would be worse due to the loss of the inter-slice contextual information, as the CNN cannot identify the inter-slice relationship \cite{zhang_automatic_2018}. It is, therefore, straightforward to generalize the conventional 2D convolution layers to 3D ones. Compared to the 2D images, the 3D volumetric data occupies much larger memory space, especially when training the network. In deep learning, the GPU has to store not only the input images, but also the activation maps of all neurons to compute the gradients during training. Therefore, when dealing with 3D volumetric images, one usually facing a pressing dilemma between memory efficiency and segmentation accuracy. The segmentation accuracy will be lost due to the absence of spatial contextual information if we treat the volumetric image slices as separate 2D images, while the training may become infeasible if we feed the whole volume into the network. One of the solutions to the dilemma is to split the whole volumetric image into 3D image patches, which is considered to be beneficial in reducing memory cost while preserving the inter-slice contextual information. In fact, it is the most commonly adopted approach in designing 3D-CNN based methods\cite{cicek_3d_2016, kamnitsas_deepmedic_2016, kamnitsas_efficient_2017, chen_voxresnet:_2018, zhang_automatic_2018, isensee_nnu-net_2018, chang_brain_2018}, and has achieved remarkable performance in various tasks such as brain lesion and tumor segmentations\cite{maier_isles_2017, menze_multimodal_2015}.

Despite that splitting into 3D patches reduces the memory cost while preserving the spatial correlations across slices, it in turn brings another spatial contextual information loss, as the fields of view (FoVs) of the CNNs were strictly limited by the patch size. When the foreground object is large, or located near the edge of a patch, the segmentation accuracy will be reduced due to the loss of semantic information. To enlarge the FOV while preserving memory feasibility,  Kamnitsas \textit{et al.} proposed to introduce downsampled images to their DeepMedic model as an auxiliary\cite{kamnitsas_deepmedic_2016}, where the output feature maps of the downsampled and the original image patches were fused before the last convolution layer to generate the final segmentation map. Owining to its multi-resolution input structure, the DeepMedic won in both ischemic stroke lesion segmentation and brain tumor segmentation challenges in 2015. Inspired by DeepMedic, we propose to adopt patches with different resolutions as inputs. To further improve the performance, the local and global features were fused in a layer-by-layer manner.

U-Net has been one of the most successful structure in medical image segmentation\cite{Ronneberger2015}. It is designed as a symmetric encoder-decoder structure, with several skip connections between the encoder and decoder to refine the segmentation results, as depicted in Fig. \ref{fig:basic_structure}a. To tackle the limit-FoV problem of 3D-volumetric segmentation networks, we propose to add another branch of encoder for the U-Net, as shown in Fig. \ref{fig:basic_structure}b. Two encoders, denoted as the local branch and the global branch, take patches with the same matrix sizes but different resolutions as inputs. In particular, the local branch uses the patches with original resolution as input to extract local patterns, while the global branch uses the downsampled patches to enlarge the FoVs. The feature maps from the local and global branches were fused at every layer and fed into the decoder network for generating the segmentation results. As we will show in this paper, the proposed double-branch residual-structured U-Net (DB-ResUNet) structure can achieve better segmentation accuracy than its single-branch counterpart.

To evaluate the segmentation performance, we collected 41 subjects from Tianjin First Central Hospital (Tianjin, China), and pretended to segment seven grey matter nuclei, including CN, GP, PUT, THA, SN, RN and DN in both left and right hemispheres. By training on 20 subjects, the proposed DB-ResUNet is able to achieve much better performance than the atlas-based method and the single-branch U-Net.

\section{Method}

\begin{figure}
    \centering
    \includegraphics[width=0.4\linewidth]{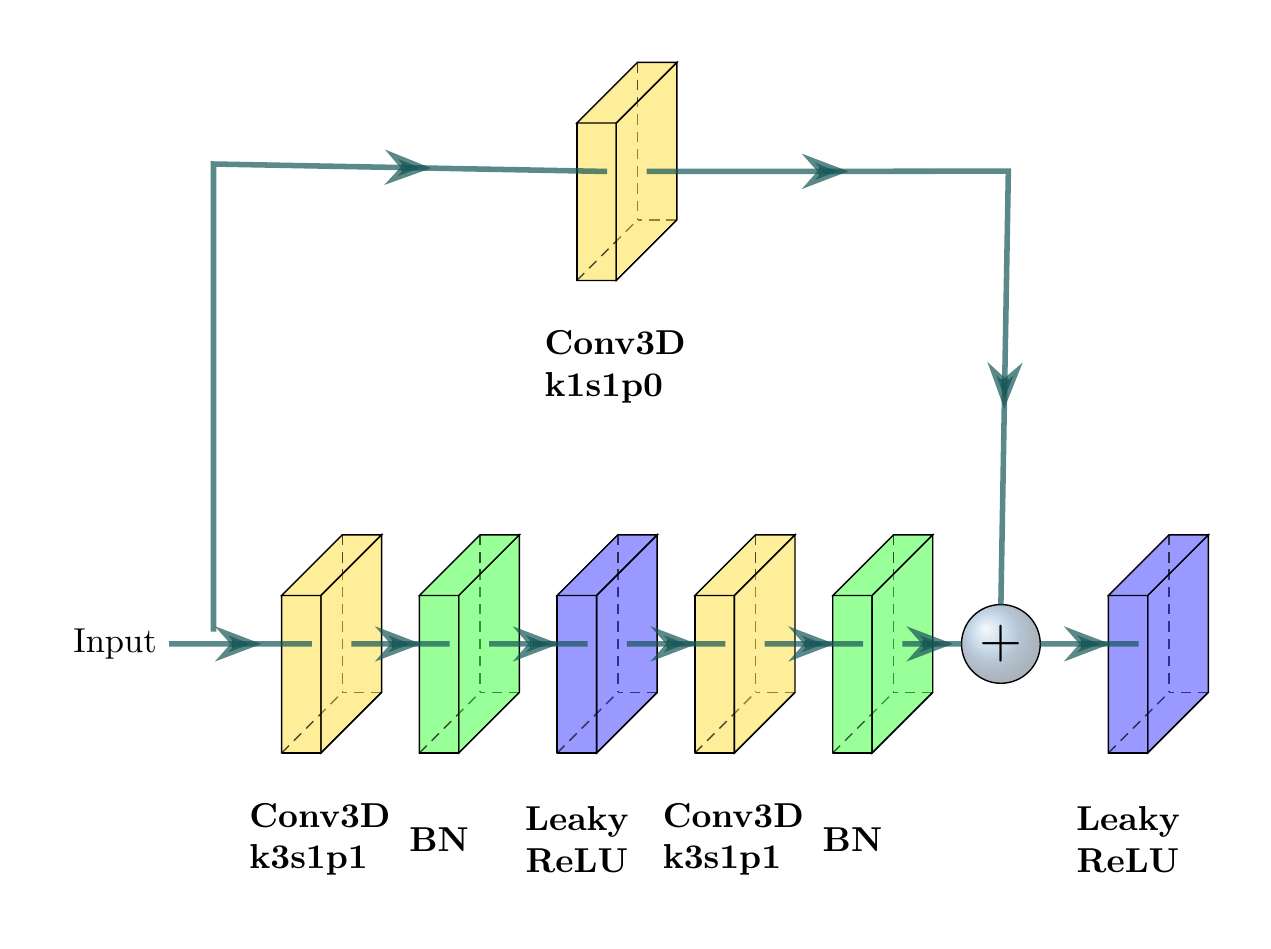}
    \caption{Architecture of a residual block. $k$, $s$ and $p$ represent the kernel size, stride and padding size, respectively.}
    \label{fig:resblock}
\end{figure}

\subsection{Residual-Structured U-Net}

\begin{figure}
    \centering
    \includegraphics[width=0.4\linewidth]{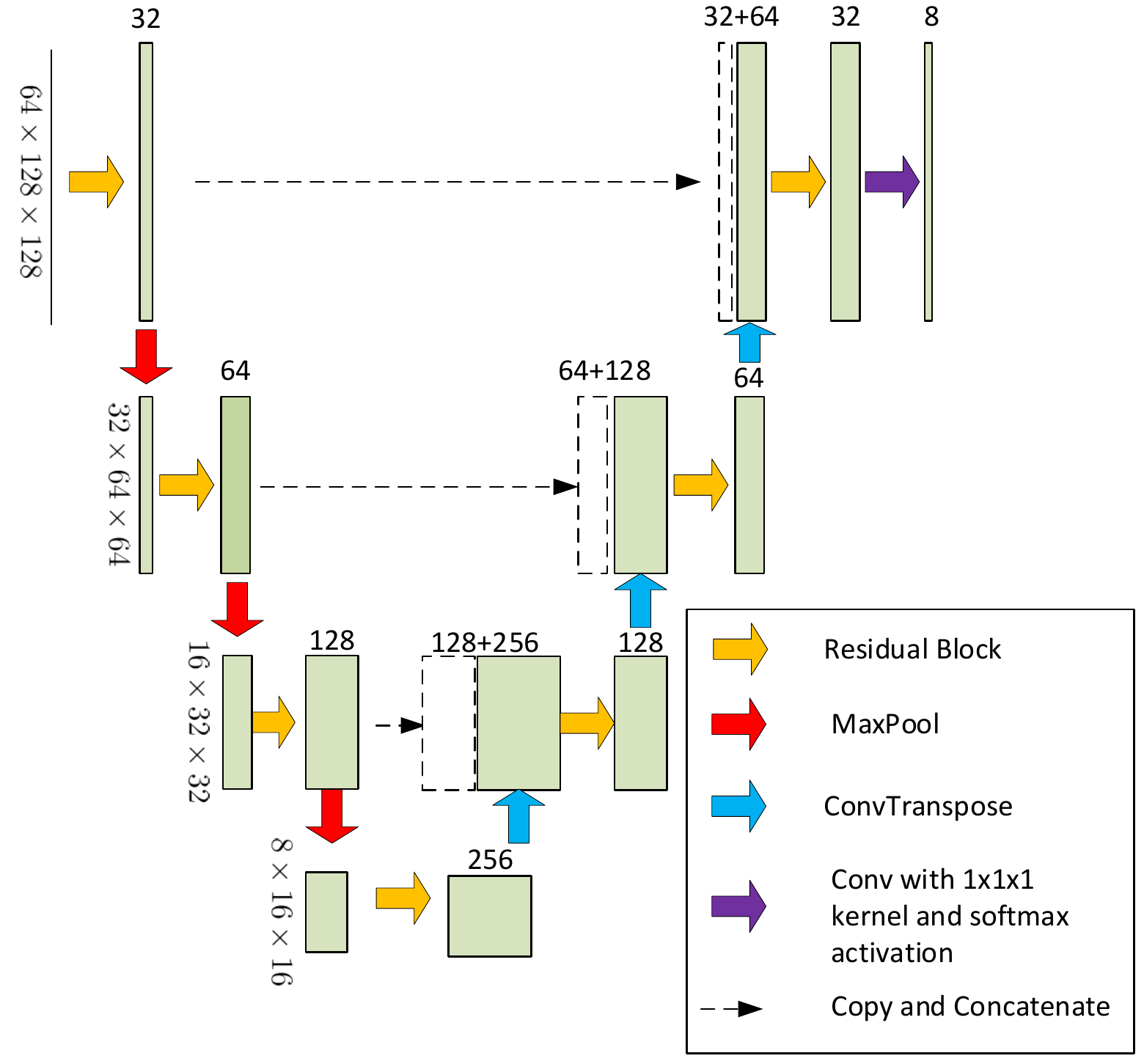}
    \caption{Architecture of ResUNet. Compared to the original U-Net, the ResUNet use residual blocks with 3D convolution layers to extract features.}
    \label{fig:resunet}
\end{figure}

In this paper, a 3D residual-structured U-Net (ResUNet) is first proposed as a baseline model for nuclei segmentation, as depicted in Fig. \ref{fig:resunet}. The ResUNet employs the similar structure to the U-Net with several modifications made to improve the performance.

First, we adapt the original U-Net, which used 2D images as input, to a 3D CNN, due to the fact that each nucleus appears across several slices, and better segmentation accuracy would be expected by utilizing the inter-slice spatial contextual information.

Second, the stacked convolution layers in the original U-Net are replaced by a residual block as shown in Fig. \ref{fig:resblock}. As pointed out in \cite{he_deep_2015}, the residual structure enabled the network parameters to be updated from the beginning, and elegantly solved the gradient vanishing problem.

Due to limited GPU memory, the images are divided into patches of size $64\times 128\times 128$ before fed into the network. Note that although feeding patches instead of the whole image has been a regular approach in 3D CNNs, only utilizing local context information from the patches may led to segmentation performance loss. To this end, we further propose to incorporate global context information without significantly increasing the number of parameters and the memory cost, which will be introduced in the next subsection in detail.

\subsection{DB-ResUNet}

Fig. \ref{fig:basic_structure}b illustrates the overall structure of our proposed double-branch Residual-structured U-Net (DB-ResUNet). We propose to segment on 3D volumetric patches due to the fact that each nucleus appears across several slices, and using 3D patches can efficiently utilize the cross-layer information. To tradeoff between the memory efficiency and a large FoV, the proposed model is basically an encoder-decoder structure that integrated both local patches, which has a higher resolution but a smaller FoV, and global patches, which is downsampled but has a larger FoV, to generate an accurate segmentation. Despite that the local and global patches have different resolutions, we propose to use patches with the same matrix size, i.e., $64\times 128\times 128$.  We employ residual block shown in Fig. \ref{fig:resblock} as the basic brick to extract features, and adopt weighted crossentropy loss function to improve its optimization.

\subsubsection{Local Branch}\label{sec:local}

\begin{figure}
    \centering
    \includegraphics[width=0.6\linewidth]{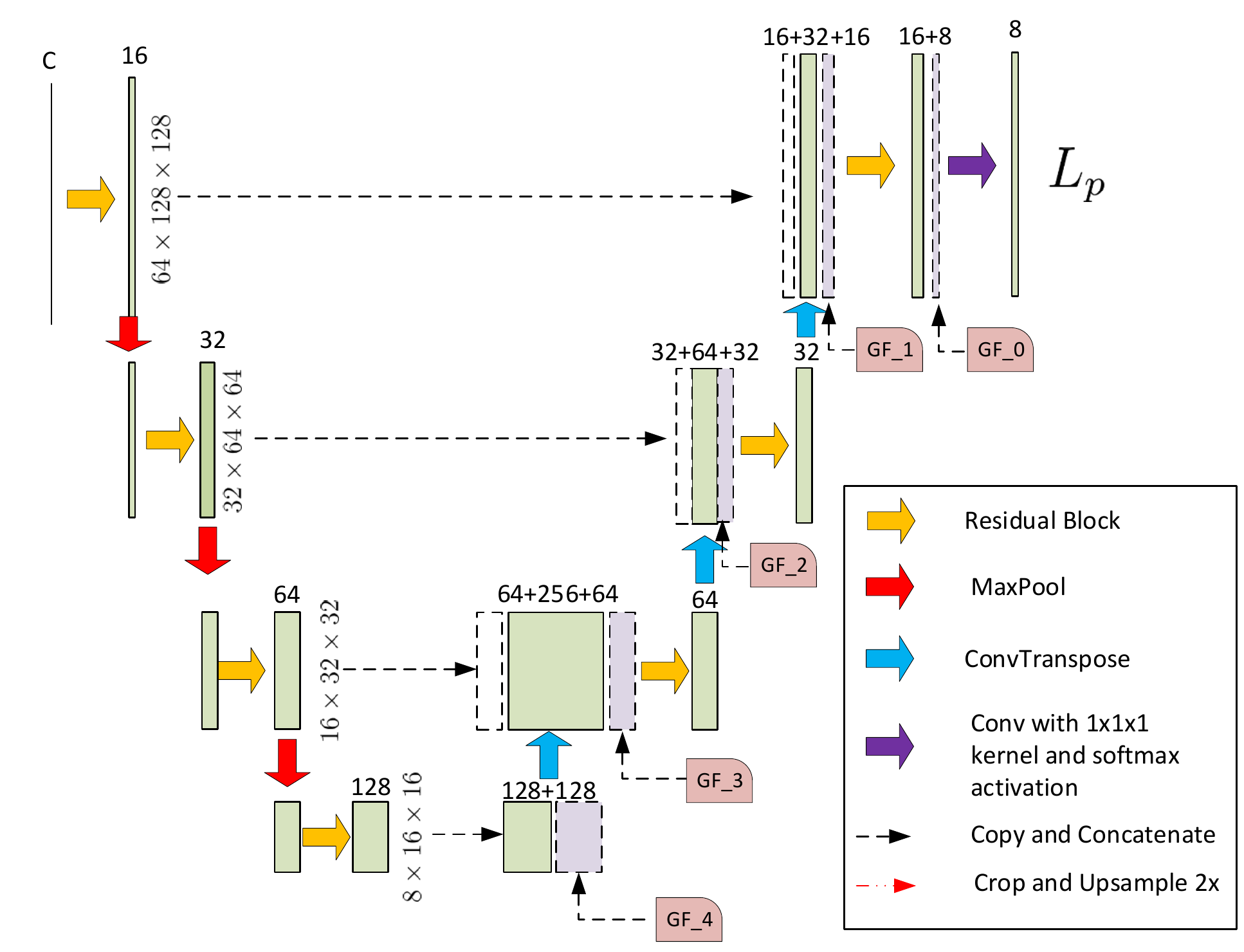}
    \caption{Architecture of the main branch of DB-ResUNet. }
    \label{fig:dbresunet_local}
\end{figure}

Fig. \ref{fig:dbresunet_local} presents the main branch of our proposed DB-ResUNet. The local patches are fed into this local branch, and the final segmentation map is generated with the assistance from the feature maps extracted from the global branch. The local branch network is basically a U-Net with residual block, which is similar to the ResUNet structure introduced in Sec. III-A.s We directly cut the images to patches of size $64\times 128\times 128$ as input, and fed into the network. The encoder part employs a similar structure with 3D UNet except for that we use residual blocks, instead of stacked convolution layers to extract features.

In the decoder part, the feature maps are upsampled by learnable convolution transpose layers. The upsampled feature maps are concatenated with the feature maps from both the local and global branch encoders, followed by a residual block. The feature maps from the global branch compensate the loss of spatial contextual information due to the limited patch size, and is beneficial in improving the segmentation accuracy.

\subsubsection{Global Branch}
\begin{figure}
    \centering
    \includegraphics[width=0.4\linewidth]{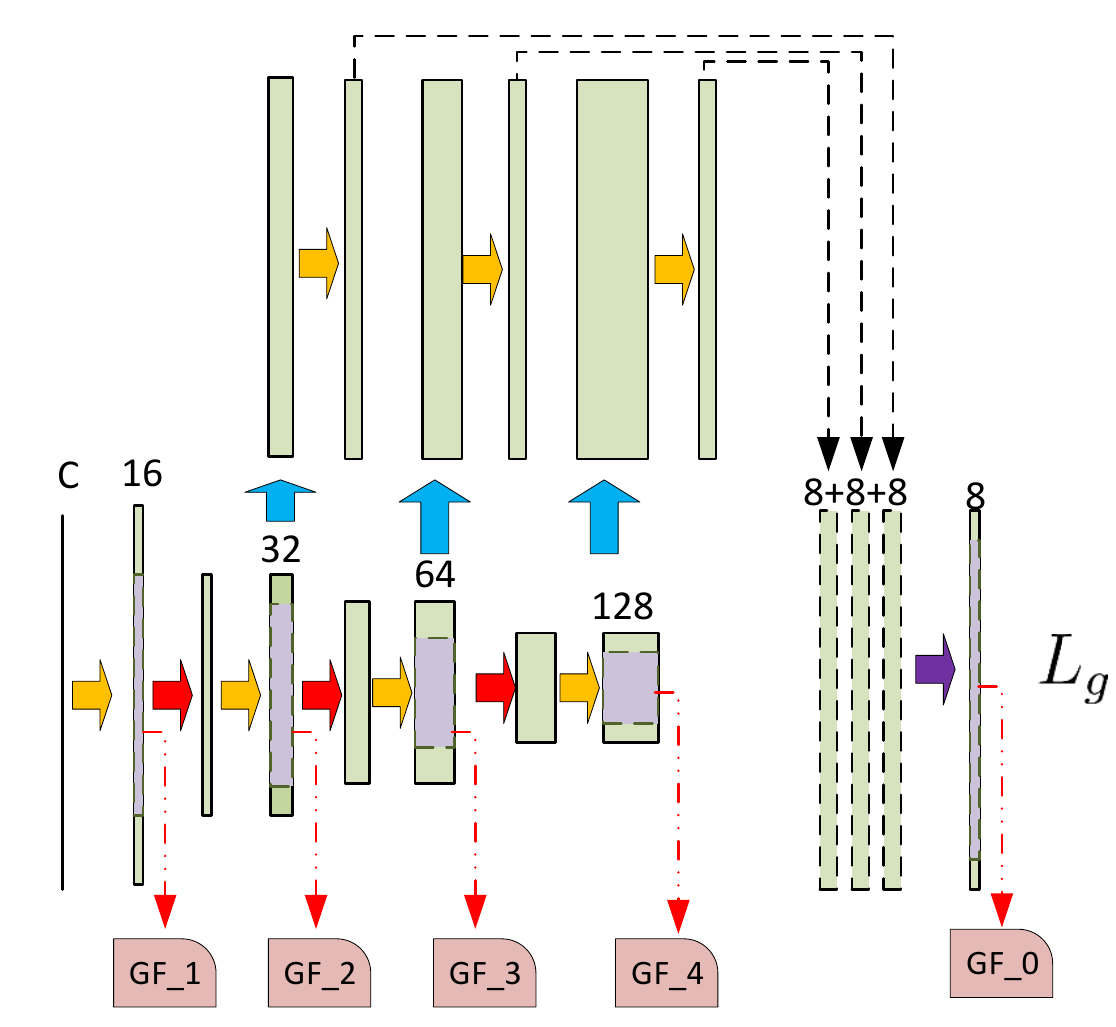}
    \caption{Architecture of global branch encoder and auxiliary decoder in DB-ResUNet.}
    \label{fig:dbresunet_global}
\end{figure}

Fig. \ref{fig:dbresunet_global} illustrates the global branch encoder. The global branch employs a fully-convolutional network (FCN) structure. The original input image was first downsampled, and then fed into the network. The output feature maps of each residual block are cropped to the same FoVs as the corresponding feature maps in the local branch, upsampled to the same spatial resolution, and then concatenated to the decoder part of the local branch.

The global branch also generates its own downsampled segmentation map by employing a FCN-8s head. The generated downsampled segmentation map is also cropped, upsampled and concatenated with the feature maps before the final convolution layer of the local branch.

\subsection{Loss Function}

Loss function is used to measure the differences between the predicted image and the label, which provides gradients to update the network parameters. In our work, we use the weighted sum of the losses from both the global and local branches as
\begin{equation}
	L(h_\theta(x),y)=L_p(h_\theta(x),y)+\lambda_g L_g(h_\theta(x),y),
\end{equation}
where $x$ and $y$ are data and the corresponding labels, respectively. $\theta$ is the network parameters, $h_\theta(x)$ denotes the predicted output of the network. $L_p(h_\theta(x),y)$ and $L_g(h_\theta(x),y)$ denotes the loss functions of the final segmentation and the global auxiliary result, respectively. $\lambda_g$ is a tradeoff constant. The global branch loss $L_g$ in fact serves as an regularization, and helps improving the segmentation accuracy.

We propose to use weighted cross entropy loss in both the global and local loss functions. By considering the fact that there are significantly more background pixels than the foreground ones, in both $L_p(h_\theta(x),y)$ and $L_g(h_\theta(x),y)$, the weights for the background and foreground categories were assigned as $0.1$ and $0.4$, respectively.

\subsection{Evaluation Metrics}

Dice coefficient (DC) is adopted to evaluate the segmentation performance. In particular, for the $i$-th category, the DC is defined as
\begin{equation}
	DC_i=\frac{2|P_i\cap G_i|}{|P_i\cup G_i|},
\end{equation}
where $P_i$ and $G_i$ denote the prediction and ground truth, respectively. $|\cdot|$ denote the area.
%

\section{Experiment Results}

\subsection{Data Acquisition and Preprocessing}

We collected 43 subjects from Tianjin First Central Hospital (Tianjin, China), where the mean age is $35.97\pm 12.11$. Ethic approval has been granted by Tianjin First Central Hospital Ethic Committee. The 3D T$_\text{1}$ weighted imaging (T$_\text{1}$WI) and the susceptibility weighted imaging (SWI) were acquired using a 3.0T MR scanner (TRIO scanner, Siemens Healthineers, Erlangen, Germany), with an 8-channel phased array head coil. The voxel sizes for 3D T$_\text{1}$WI and SWI are $1\mathrm{mm}\times 1\mathrm{mm}\times 1\mathrm{mm} $ and  $2\mathrm{mm}\times 0.5134\mathrm{mm}\times 0.5134\mathrm{mm} $, respectively. The corresponding matrix sizes for 3D T$_\text{1}$WI and SWI are $176\times 256\times 256$ and $56\times 336\times 448$, respectively. The magnitude and phase images of SWI were used to generate the QSM following the method in \cite{Chen2018}.

As shown in Fig. \ref{fig:nuclei}, in this paper, we focus on the segmentation of gray matter nuclei, including CN, GP, PUT, THA, SN, RN and DN, which were manually annotated by trained neuroradiologists (Dr. Chao Chai with 9 years' experience in neuroimaging). The whole dataset was randomly split as training set, validation set and test set, with 20, 4 and 19 subjects, respectively. The training and validation sets were used for tuning the network parameters and hyperparameters. The test set was used for evaluating the segmentation performance only.

Note that the T$_\text{1}$WI and the SWI were different in both spatial resolution and matrix size. The T$_\text{1}$WI was first registered to the corresponding SWI by adopting rigid affine transform with the mutual information as the criterion, and then resampled to the same spatial resolution and matrix size as the SWI using linear interpolator. Both the QSM and the T$_\text{1}$WI were then zero-padded to matrix sizes of $64\times 336\times 448$ before use. Finally, we clip the intensities on the QSM and the T$_\text{1}$WI to the range of $[-150, 250]$ and $[0,800]$, respectively, and linearly rescale them to the range of $[0,1]$. The T$_\text{1}$WI and the QSM are then concatenated as dual-channel images before fed into the network.

\subsection{Implementation Setup}\label{sec:implimentation}

The experiments were performed on a workstation with an Intel Core i7-7700K CPU, 64GB RAM and Nvidia GeForce 1080Ti GPU with 11GB memory. The workstation operated on Linux (Ubuntu 14.04 LTS) with CUDA 8.0. The network was implemented on PyTorch v1.0\cite{Paszke2017}. The MR image files were stored as Neuroimaging Informatics Technology Initiative (NIfTI) format, and processed using Simple Insight Toolkit (SimpleITK)\cite{Lowekamp2013}. The visualized results were presented by using ITK-SNAP\cite{Yushkevich2006}.

Data augmentation methods, including random flipping along three axes and random rotation around z-axis, were adopted to prevent overfitting, where the rotation were restricted on the $x-y$ plane within an angle range of $[-30^\circ, 30^\circ]$. We adopted Adam method\cite{Kingma2014} as the optimizer, and set the initial learning rate as $3\times 10^{-4}$ with $\beta_1=0.9$ and $\beta_2=0.99$, with a weight decay of $3\times 10^{-5}$. Due to limited memory space on a single GPU, the batch size was set to be $4$. The learning rate is reduced by a factor of $\sqrt{0.1}$ if no progress was observed on the validation loss, and the training was terminated if the learning rate was smaller than $10^{-6}$. The tradeoff coefficient $\lambda_g$ was set to be 1.

\subsection{Experiment Results}\label{sec:seg_qsm_t1}

\begin{table*}[h!]
    \caption{Mean Dice coefficient on the test set. The inference results of MRICloud is obtained by uploading the data to their website. The most outstanding result of each column has been highlighted in bold. }
    \label{tab:qsm t1 dice}
    \centering
    \begin{tabular}{c|ccccccc}
    \toprule[2pt]
     & CN & GP & PUT & THA & SN & RN & DN \\
     \midrule[1pt]
     3D-UNet& 0.788& 0.824 & 0.810 & 0.833 & 0.703 & 0.734 & 0.744\\
     ResUNet& 0.794& 0.826 & 0.804 & \textbf{0.853} & 0.704 & 0.741 & 0.749\\
     DB-ResUNet& \textbf{0.802} & \textbf{0.840} & \textbf{0.827} & 0.844& \textbf{0.719} & \textbf{0.763} & \textbf{0.774}\\
     MRICloud & 0.589 & 0.795& 0.724 & 0.692 & 0.520 & 0.697& 0.639\\
     \bottomrule[2pt]
    \end{tabular}
\end{table*}

\begin{table*}[h!]
    \centering
    \caption{Comparison of the number of parameters and the inference time of 3D-UNet, ResUNet, and DB-ResUNet. The inference time is evaluated on a workstation with Intel Core i7-7700K with 64GB RAM and a single nVidia 1080Ti GPU. The inference time of MRICloud is obtained from \cite{li_multi-atlas_2019}. }
    \label{tab:para}
    \begin{tabular}{c|ccc}
    \toprule[2pt]
        Method & \# of parameters & Inference Time (GPU) & Inference Time (CPU) \\
        \midrule[1pt]
         3D-UNet& $26$ M & $1.273\pm0.020 $s& $1868.33\pm 3.43$s\\
         ResUNet& $5.7$M & $1.305 \pm 0.033$s& $1860.68\pm 2.73$s \\
         DB-ResUNet & $4.5$M & $1.293\pm 0.056$s& $1863.16\pm 7.54$s \\
         MRICloud & / & /&$1.6\pm 0.5$h\cite{li_multi-atlas_2019}\\
         \bottomrule[2pt]
    \end{tabular}
\end{table*}

\begin{figure}
    \centering
    \includegraphics[width=.7\linewidth]{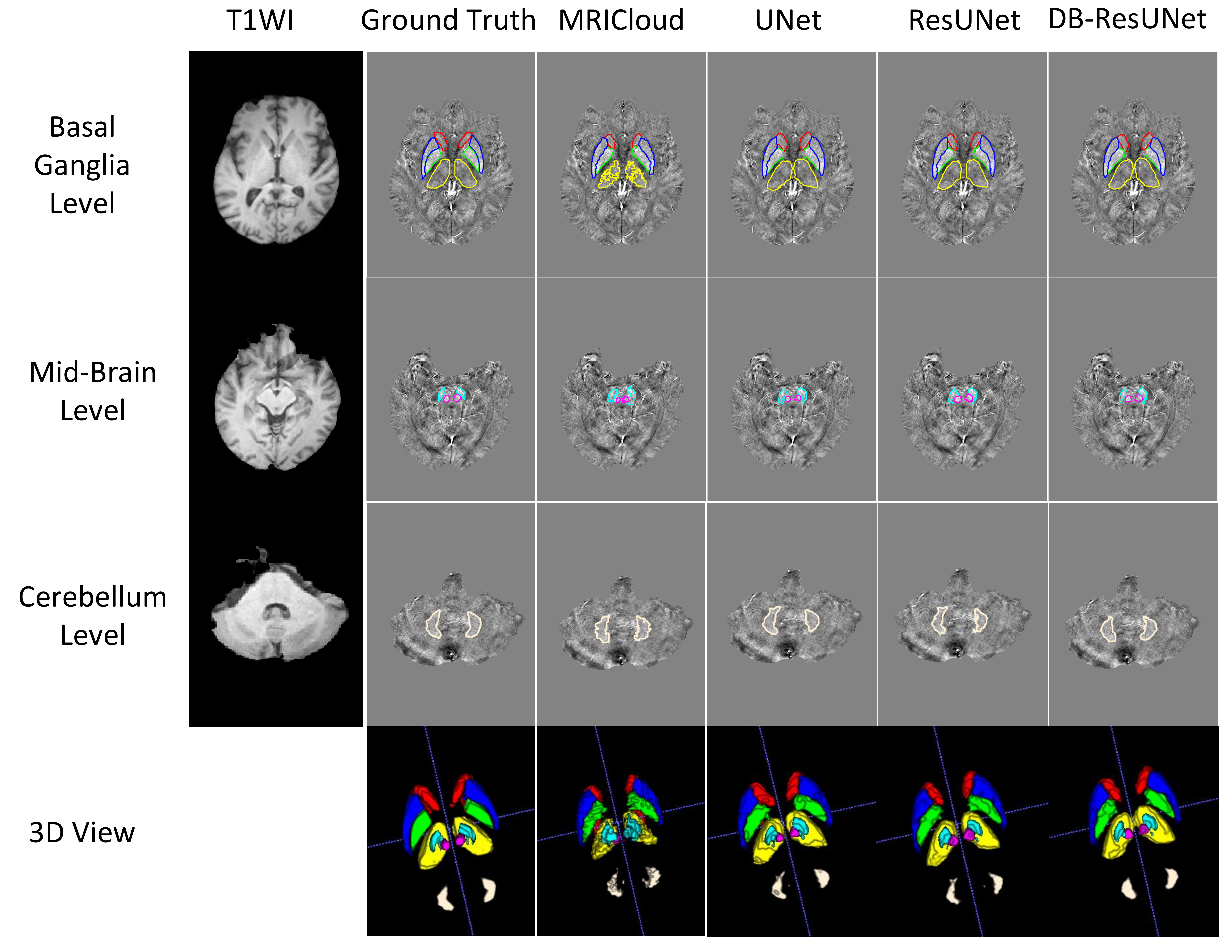}
    \caption{Visualized example of segmentation results with various automatic segmentation methods using combined T$_\text{1}$WI and QSM as input. }
    \label{fig:qsm_t1_example}
\end{figure}

Fig. \ref{fig:qsm_t1_example} presented some examples of the predicted segmentation maps of the proposed ResUNet and DB-ResUNet. For comparison, the segmentation results of an atlas-based method \cite{li_multi-atlas_2019} and another deep-learning-based method, i.e., 3D-UNet\cite{cicek_3d_2016}, were also presented. The 3D-UNet was trained on the same training samples with the same size of patches as ResUNet and DB-ResUNet. The results of the atlas-based method \cite{li_multi-atlas_2019} were obtained by uploading our test data samples to their website\footnote{http://www.mricloud.org}.

As we can see from Fig. \ref{fig:qsm_t1_example}, the segmentation maps produced by the atlas-based method were less smooth and with many false positives. The deep-learning methods, on the other hand, were able to extract features from various levels, leading to much smoother segmentations, which can be clearly observed from their 3D views presented in the last row of Fig. \ref{fig:qsm_t1_example}.
%

The numerical results on the test set were further summarized in Tab. \ref{tab:qsm t1 dice}. As we can see, the deep learning methods were in general much better than the MRICloud result. Basically, it is required in the atlas-based method to align the images to the MNI template, so that the nuclei segmentations can be performed by comparing the registered image and the images in the training set in a pixel-by-pixel manner. Clearly, the atlas-based method has been more vulnerable to the deformation of brain structure. The deep-learning based methods, however, enabled the neural network to extract multi-level features by itself, making it more robust to the image deformation, leading to a better segmentation performance.

We can also observe from Tab. \ref{tab:qsm t1 dice} that thanks to the residual structure, the ResUNet achieved better segmentation accuracy than the 3D-UNet. In fact, it was shown in \cite{hardt_identity_2018} that the use of residual structures makes the objective function more smooth, making the convergence point closer to the global optima. The DB-ResUNet, on the other hand, achieved the best segmentation performance. The improvement should contribute to the larger FoV brought by the global branch. With a larger FoV, the network is able to incorporate more spatial context information in segmentation, making the voxel-wise classification more accurate.

\begin{figure*}
    \centering
    \includegraphics[width=.8\linewidth]{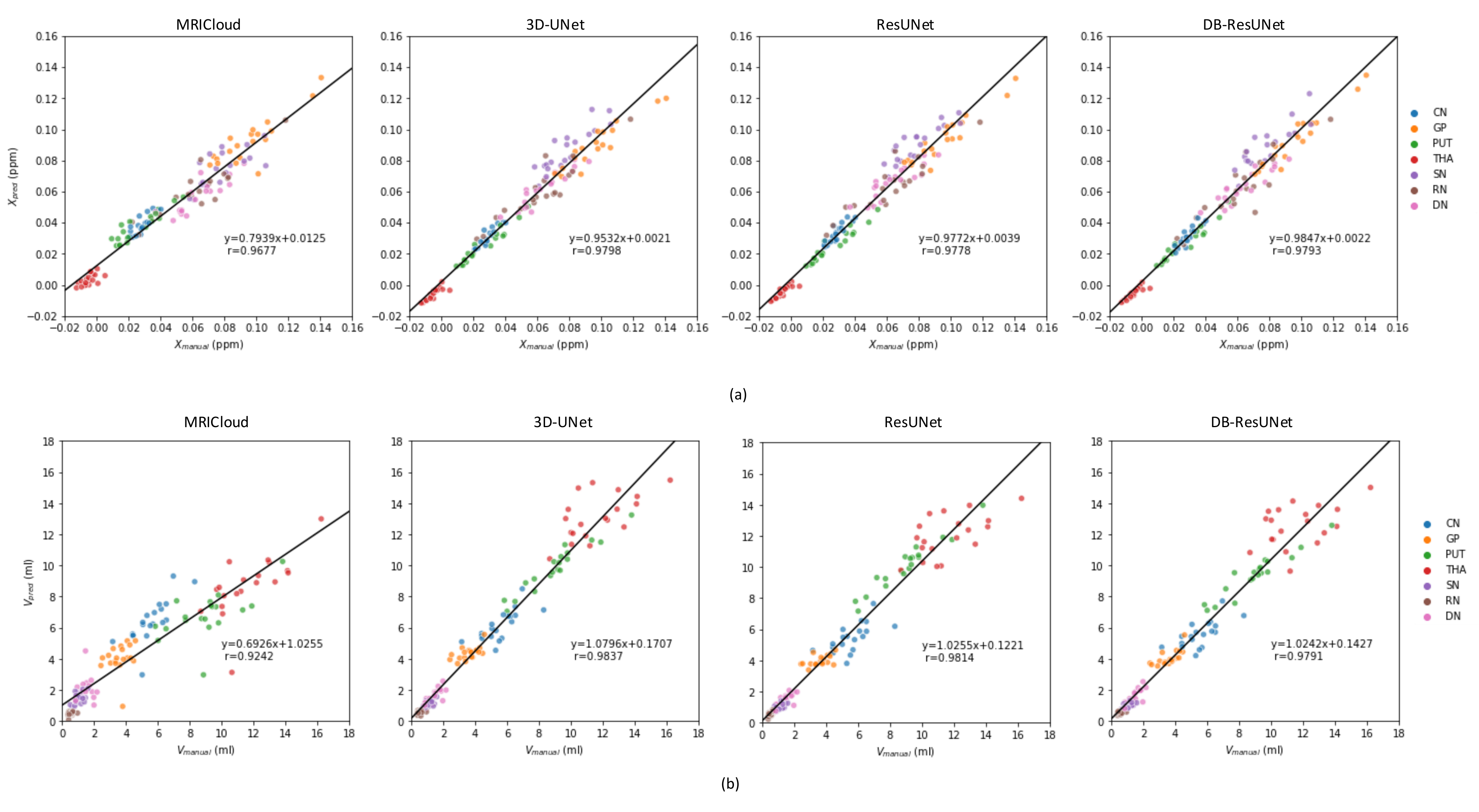}
    \caption{Scatter plots between various predicted segmentation using QSM and T$_\text{1}$WI as input and the manual delineation. (a) Susceptibility values. (b) Volumes.}
    \label{fig:scatter_methods}
\end{figure*}

Fig. \ref{fig:scatter_methods} further compared the susceptibility values and the volumes between the predicted and manual segmentations. The regression line was also plotted. As we can see, deep-learning-based methods were more accuracy in both susceptibility values and volumes, where the DB-ResUNet achieves the best performance, which highlights the outstanding performance of the proposed method.

Another advantage of deep-learning-based methods lied in its fast inference. As reported in \cite{li_multi-atlas_2019}, the method adopted in MRICloud take $1.6\pm 0.5$h to process one subject. Deep learning methods, as summarized in Tab. \ref{tab:para}, consumed much shorter time, where the proposed DB-ResUNet achieved an average time of $1.293\pm 0.056$s to generate segmentation map for each subject. To make a fair comparison, we also evaluated the results by running the trained network on the CPU (Intel Core i7-7700K with 64GB RAM). As Tab. \ref{tab:para} shows, the CPU running time, despite significantly longer than the GPU running time, is also much shorter than the atlas-based method, which highlights the computational advantage by using deep learning.

The numbers of parameters of the deep-learning methods are also summarized in Tab. \ref{tab:para}. As we can see, by replacing the stacked convolution layers by a residual block, the ResUNet was able to achieve better segmentation performance, while at the same time significantly reduces the number of parameters. Also note from Figs. \ref{fig:resunet} and \ref{fig:dbresunet_local} that the numbers of filters in both global and local branches in the DB-ResUNet were halved compared to the ResUNet. Therefore, despite that an additional branch was added, the number of parameters can be further reduced, which brings a shorter computation time during inference.

\section{Discussion}

\subsection{Impact of Downsampling Rate}

In the global branch of the proposed DB-ResUNet, the image patches were cropped from the image that downsampled for 2 times, i.e., the width and height were halved compared to the original image. Note that compared to the baseline model, i.e., ResUNet, DB-ResUNet also introduces an auxiliary output for the global branch, and the auxiliary loss is added as a regularizer. It is, therefore, necessary to discuss whether the performance gain of DB-ResUNet really comes from the enlarged FoV.

Tab. \ref{tab:downsample} presented the numerical results where the global branch input patches were cropped from images downsampled by different rates. As we can see, the 2x case achieves the best segmentation accuracy, while the 4x case achieves worse performance in small nuclei, such as SN, RN and DN. Intuitively, with a larger downsampling rate, the global branch input brings more spatial contextual but less detailed information, leading to worse segmentation accuracy in small nuclei.

\begin{table}[]
    \centering
    \caption{Mean Dice coefficient of DB-ResUNet on the test set with different scale of global branch input.}
    \label{tab:downsample}
    \begin{tabular}{c|ccccccc}
        \toprule[2pt]
         Rate & CN & GP & PUT & THA & SN & RN & DN \\
         \midrule[1pt]
         1x& 0.731 & 0.779 & 0.755 & 0.822& 0.679 & 0.687 & 0.490\\
         2x& 0.802 & 0.840 & 0.827 & 0.844& 0.719 & 0.763 & 0.774\\
         4x& 0.530 & 0.736 & 0.733 & 0.772 & 0.087 & 0.212 & 0.083  \\

         \bottomrule[2pt]
    \end{tabular}
\end{table}

\subsection{Segmentation on Either T$_\text{1}$WI or QSM}

\begin{figure}
    \centering
    \includegraphics[width=0.7\linewidth]{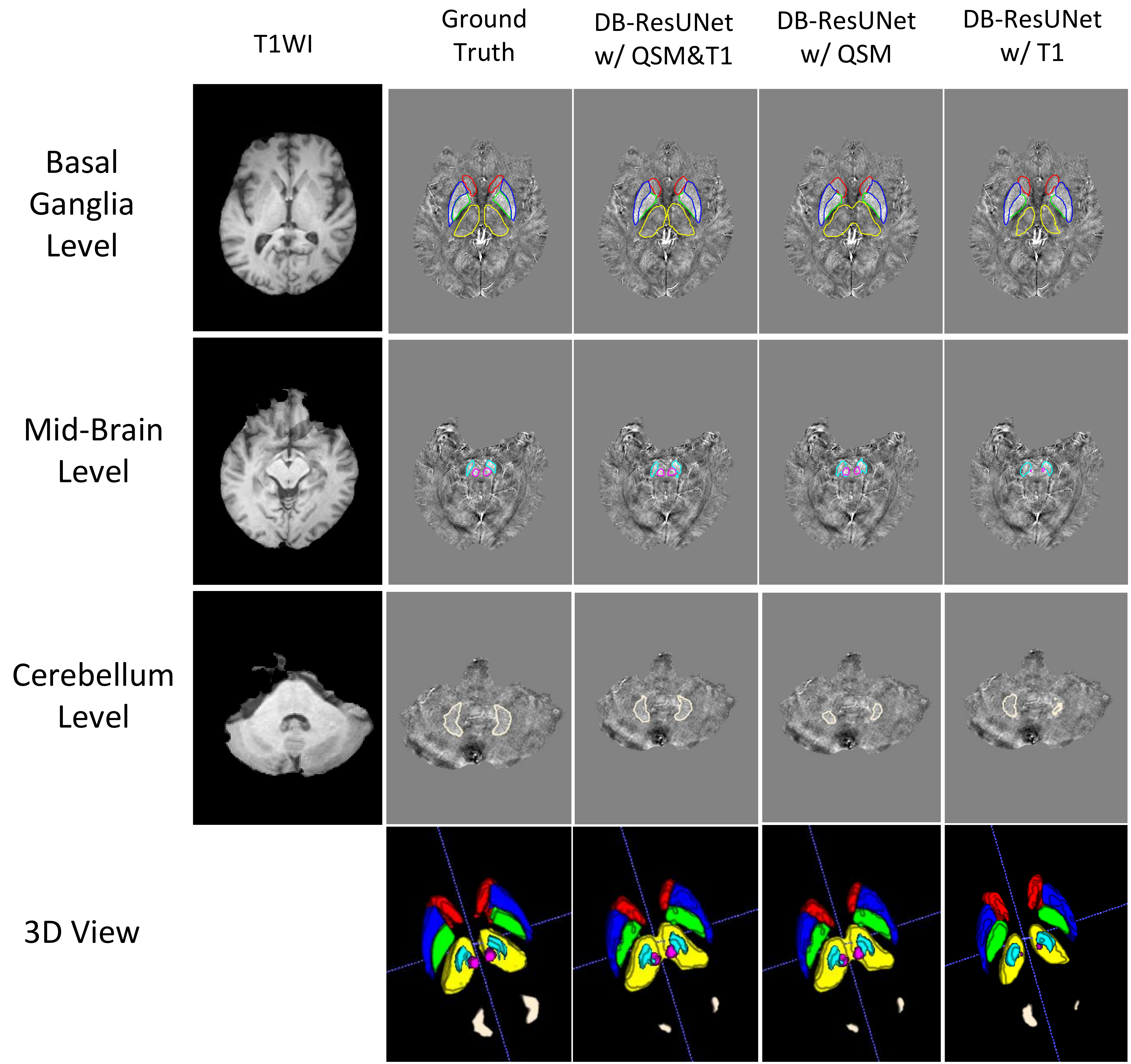}
    \caption{Visualized example of segmentation results with the proposed DB-ResUNet with various inputs. The segmentation results are overlaid on the QSM.}
    \label{fig:dbunet_qsm_or_t1}
\end{figure}

\begin{figure}
    \centering
    \includegraphics[width=.5\linewidth]{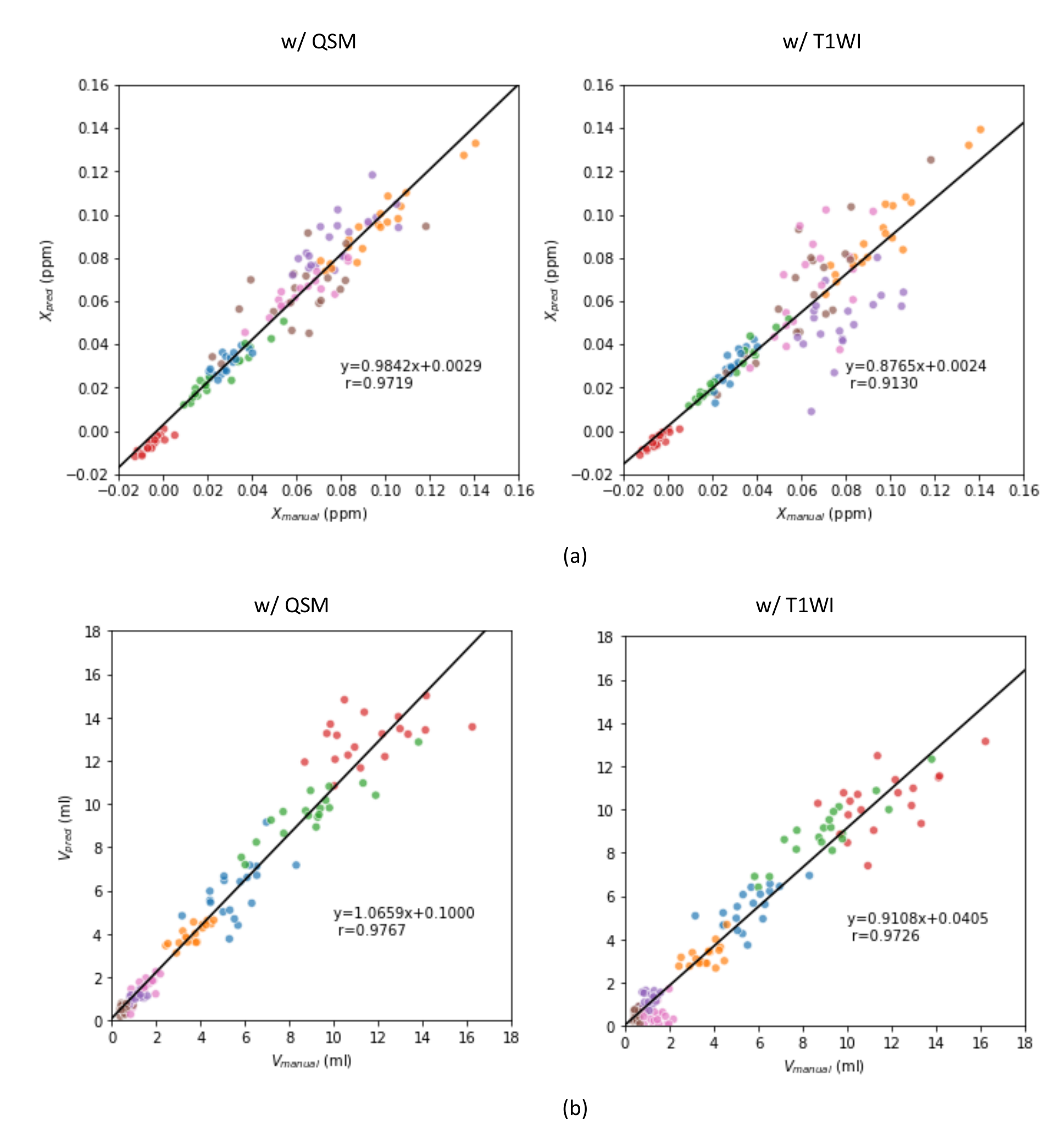}
    \caption{Scatter plots between DB-ResUNet predicted segmentation using either QSM or T$_\text{1}$WI as input and the manual delineation. (a) Suspectibility values. (b) Volumes.}
    \label{fig:scatter_inputs}
\end{figure}

It has been shown in Sec. \ref{sec:seg_qsm_t1} that the proposed segmentation method achieved high segmentation accuracy. One limitation of the T$_\text{1}$WI/QSM-based method was that we need to obtain both T$_\text{1}$WI and QSM images, which may not be feasible in all studies. Therefore, we will further assess the performance of the proposed DB-ResUNet when only QSM or T$_\text{1}$WI is available.

In particular, we adopted either the T$_\text{1}$WI or the QSM as a single-channel image, and fed it into the network. Fig. \ref{fig:dbunet_qsm_or_t1} presented the visualized example of the segmentation results. For the sake of comparison, the segmentation results by using both the T$_\text{1}$WI and QSM were also plotted. As we can see from Fig. \ref{fig:dbunet_qsm_or_t1}, merely using one image may lead to worse performance.

Tab. \ref{tab:qsm or t1} presented the numerical evaluation results. As we can see, with only T$_\text{1}$WI, the method achieved the worst segmentation accuracy due to the low T1 contrast in the iron-rich deep gray matter nuclei, which was also confirmed by the comparison between the predicted and manual annotated results in both susceptibility values and volumes shown in Fig. \ref{fig:scatter_inputs}. In fact, the QSM, despite of its low structural resolution, has advantage in presenting the iron-accumulating deep gray matter nuclei. Our study found that the segmentation accuracy was the highest based on the combined QSM/T$_\text{1}$ images compared with that based on the only QSM or T$_\text{1}$WI images, and we also found that the segmentation accuracy was higher based on the only QSM images compared with that based on the only T$_\text{1}$WI. In the atlas-based nuclei segmentation methods, most pipelines \cite{igual_fully-automatic_2011,igual_automatic_2012,su_thalamus_2019} compulsorily adopted the T$_\text{1}$WI for segmentation, so that an accurate registration between the subject-in-question's and the MNI template can be achieved. Our method, other other hand, suggested that with deep learning, a reasonably accurate nuclei segmentation result can be achieved, even when only QSM is available.

\begin{table}[h!]
    \caption{Mean Dice coefficient of DB-ResUNet on the test set with different input data. The segmentation results that use both QSM and T$_\text{1}$WI are also presented for comparison.}
    \label{tab:qsm or t1}
    \centering
    \begin{tabular}{c|ccccccc}
   \toprule[2pt]
   Input & CN & GP & PUT & THA & SN & RN & DN \\
     \midrule[1pt]
     QSM & 0.782&0.828 &0.819 &0.827&0.690&0.680&0.714 \\
     T$_\text{1}$WI & 0.760 & 0.778 & 0.799 & 0.825&
     0.589 & 0.635 & 0.338 \\
     \begin{tabular}{c}
     QSM\\+T$_\text{1}$WI
     \end{tabular} & 0.802 & 0.840 & 0.827 & 0.844& 0.719 & 0.763 & 0.774\\
     \bottomrule[2pt]
    \end{tabular}
\end{table}

\section{Conclusion}

In this paper, we proposed to adopt a 3D CNN-based method to segment gray matter nuclei from the QSM and T$_\text{1}$WI. To tackle the loss of FoVs that comes from splitting 3D volumetric images into patches, we proposed to adopt a double-branch network that takes patches with both original and low resolution as input. The proposed method achieved much higher segmentation accuracy than either the atlas-based method or the conventional 3D-UNet. The contribution of T$_\text{1}$WI and QSM was also discussed. Experimental results implies that the QSM had more contribution in distinguishing the nuclei, which made it possible for accurate quantitative susceptibility assessment when 3D T$_\text{1}$WI was absent.

\bibliographystyle{ieeetr}
\bibliography{main.bib}

\begin{thebibliography}{10}

\bibitem{li_association_2015}
W.~Li, C.~Langkammer, Y.-H. Chou, K.~Petrovic, R.~Schmidt, A.~W. Song, D.~J.
  Madden, S.~Ropele, and C.~Liu, ``Association between increased magnetic
  susceptibility of deep gray matter nuclei and decreased motor function in
  healthy adults,'' {\em Neuroimage}, vol.~105, pp.~45--52, Jan. 2015.

\bibitem{Chai2019}
C.~Chai, H.~Wang, S.~Liu, Z.-Q. Chu, J.~Li, T.~Qian, E.~Haacke, S.~Xia, and
  W.~Shen, ``Increased iron deposition of deep cerebral gray matter structures
  in hemodialysis patients: A longitudinal study using quantitative
  susceptibility mapping,'' {\em Journal of Magnetic Resonance Imaging},
  vol.~49, no.~3, pp.~786--799, 2019.

\bibitem{Gong2019}
N.-J. Gong, R.~Dibb, M.~Bulk, L.~{van der Weerd}, and C.~Liu, ``Imaging beta
  amyloid aggregation and iron accumulation in alzheimer's disease using
  quantitative susceptibility mapping mri,'' {\em NeuroImage}, vol.~191,
  pp.~176 -- 185, 2019.

\bibitem{Xuan2017}
M.~Xuan, X.~Guan, Q.~Gu, Z.~Shen, X.~Yu, T.~Qiu, X.~Luo, R.~Song, Y.~Jiaerken,
  X.~Xu, P.~Huang, W.~Luo, and M.~Zhang, ``Different iron deposition patterns
  in early- and middle-late-onset parkinson's disease,'' {\em Parkinsonism and
  Related Disorders}, vol.~44, pp.~23 -- 27, 2017.

\bibitem{Chen2019}
L.~Chen, J.~Hua, C.~A. Ross, S.~Cai, P.~C. van Zijl, and X.~Li, ``Altered brain
  iron content and deposition rate in huntington's disease as indicated by
  quantitative susceptibility mri,'' {\em Journal of Neuroscience Research},
  vol.~97, no.~4, pp.~467--479, 2019.

\bibitem{florio_basal_2018}
T.~M. Florio, E.~Scarnati, I.~Rosa, D.~Di~Censo, B.~Ranieri, A.~Cimini,
  A.~Galante, and M.~Alecci, ``The {Basal} {Ganglia}: {More} than just a
  switching device,'' {\em CNS Neurosci Ther}, vol.~24, no.~8, pp.~677--684,
  2018.

\bibitem{liu_susceptibility-weighted_2015}
C.~Liu, W.~Li, K.~A. Tong, K.~W. Yeom, and S.~Kuzminski,
  ``Susceptibility-weighted imaging and quantitative susceptibility mapping in
  the brain,'' {\em Journal of Magnetic Resonance Imaging}, vol.~42, no.~1,
  pp.~23--41, 2015.
\newblock \_eprint: https://onlinelibrary.wiley.com/doi/pdf/10.1002/jmri.24768.

\bibitem{igual_fully-automatic_2011}
L.~Igual, J.~C. Soliva, A.~Hernández-Vela, S.~Escalera, X.~Jiménez,
  O.~Vilarroya, and P.~Radeva, ``A fully-automatic caudate nucleus segmentation
  of brain {MRI}: {Application} in volumetric analysis of pediatric
  attention-deficit/hyperactivity disorder,'' {\em BioMed Eng OnLine}, vol.~10,
  p.~105, Dec. 2011.

\bibitem{igual_automatic_2012}
L.~Igual, J.~C. Soliva, S.~Escalera, R.~Gimeno, O.~Vilarroya, and P.~Radeva,
  ``Automatic brain caudate nuclei segmentation and classification in
  diagnostic of {Attention}-{Deficit}/{Hyperactivity} {Disorder},'' {\em
  Computerized Medical Imaging and Graphics}, vol.~36, pp.~591--600, Dec. 2012.

\bibitem{xia_automatic_2007}
Y.~Xia, K.~Bettinger, L.~Shen, and A.~L. Reiss, ``Automatic {Segmentation} of
  the {Caudate} {Nucleus} {From} {Human} {Brain} {MR} {Images},'' {\em IEEE
  Transactions on Medical Imaging}, vol.~26, pp.~509--517, Apr. 2007.
\newblock Conference Name: IEEE Transactions on Medical Imaging.

\bibitem{su_thalamus_2019}
J.~H. Su, F.~T. Thomas, W.~S. Kasoff, T.~Tourdias, E.~Y. Choi, B.~K. Rutt, and
  M.~Saranathan, ``Thalamus {Optimized} {Multi} {Atlas} {Segmentation}
  ({THOMAS}): fast, fully automated segmentation of thalamic nuclei from
  structural {MRI},'' {\em NeuroImage}, vol.~194, pp.~272--282, July 2019.

\bibitem{li_multi-atlas_2019}
X.~Li, L.~Chen, K.~Kutten, C.~Ceritoglu, Y.~Li, N.~Kang, J.~T. Hsu, Y.~Qiao,
  H.~Wei, C.~Liu, M.~I. Miller, S.~Mori, D.~M. Yousem, P.~C.~M. van Zijl, and
  A.~V. Faria, ``Multi-atlas tool for automated segmentation of brain gray
  matter nuclei and quantification of their magnetic susceptibility,'' {\em
  NeuroImage}, vol.~191, pp.~337--349, May 2019.

\bibitem{lecun_deep_2015}
Y.~LeCun, Y.~Bengio, and G.~Hinton, ``Deep learning,'' {\em Nature}, vol.~521,
  pp.~436--444, May 2015.

\bibitem{he_deep_2015}
K.~He, X.~Zhang, S.~Ren, and J.~Sun, ``Deep {Residual} {Learning} for {Image}
  {Recognition},'' {\em arXiv:1512.03385 [cs]}, Dec. 2015.
\newblock arXiv: 1512.03385.

\bibitem{hardt_identity_2018}
M.~Hardt and T.~Ma, ``Identity {Matters} in {Deep} {Learning},'' {\em
  arXiv:1611.04231 [cs, stat]}, July 2018.
\newblock arXiv: 1611.04231.

\bibitem{chen_deeplab_2017}
L.-C. Chen, G.~Papandreou, I.~Kokkinos, K.~Murphy, and A.~L. Yuille,
  ``{DeepLab}: {Semantic} {Image} {Segmentation} with {Deep} {Convolutional}
  {Nets}, {Atrous} {Convolution}, and {Fully} {Connected} {CRFs},'' {\em
  arXiv:1606.00915 [cs]}, May 2017.
\newblock arXiv: 1606.00915.

\bibitem{chen_encoder-decoder_2018}
L.-C. Chen, Y.~Zhu, G.~Papandreou, F.~Schroff, and H.~Adam, ``Encoder-{Decoder}
  with {Atrous} {Separable} {Convolution} for {Semantic} {Image}
  {Segmentation},'' {\em arXiv:1802.02611 [cs]}, Aug. 2018.
\newblock arXiv: 1802.02611.

\bibitem{ronneberger_u-net_2015}
O.~Ronneberger, P.~Fischer, and T.~Brox, ``U-{Net}: {Convolutional} {Networks}
  for {Biomedical} {Image} {Segmentation},'' {\em arXiv:1505.04597 [cs]}, May
  2015.
\newblock arXiv: 1505.04597.

\bibitem{isensee_nnu-net_2018}
F.~Isensee, J.~Petersen, A.~Klein, D.~Zimmerer, P.~F. Jaeger, S.~Kohl,
  J.~Wasserthal, G.~Koehler, T.~Norajitra, S.~Wirkert, and K.~H. Maier-Hein,
  ``{nnU}-{Net}: {Self}-adapting {Framework} for {U}-{Net}-{Based} {Medical}
  {Image} {Segmentation},'' {\em arXiv:1809.10486 [cs]}, Sept. 2018.
\newblock arXiv: 1809.10486.

\bibitem{dolz_ivd-net_2019}
J.~Dolz, C.~Desrosiers, and I.~Ben~Ayed, ``{IVD}-{Net}: {Intervertebral} {Disc}
  {Localization} and {Segmentation} in {MRI} with a {Multi}-modal {UNet},'' in
  {\em Computational {Methods} and {Clinical} {Applications} for {Spine}
  {Imaging}} (G.~Zheng, D.~Belavy, Y.~Cai, and S.~Li, eds.), Lecture {Notes} in
  {Computer} {Science}, (Cham), pp.~130--143, Springer International
  Publishing, 2019.

\bibitem{zhang_automatic_2018}
R.~Zhang, L.~Zhao, W.~Lou, J.~M. Abrigo, V.~C. Mok, W.~C. Chu, D.~Wang, and
  L.~Shi, ``Automatic {Segmentation} of {Acute} {Ischemic} {Stroke} from {DWI}
  using {3D} {Fully} {Convolutional} {DenseNets},'' {\em IEEE Transactions on
  Medical Imaging}, vol.~PP, no.~99, pp.~1--1, 2018.

\bibitem{cicek_3d_2016}
O.~Cicek, A.~Abdulkadir, S.~S. Lienkamp, T.~Brox, and O.~Ronneberger, ``{3D}
  {U}-{Net}: {Learning} {Dense} {Volumetric} {Segmentation} from {Sparse}
  {Annotation},'' June 2016.

\bibitem{kamnitsas_deepmedic_2016}
K.~Kamnitsas, E.~Ferrante, S.~Parisot, C.~Ledig, A.~V. Nori, A.~Criminisi,
  D.~Rueckert, and B.~Glocker, ``{DeepMedic} for {Brain} {Tumor}
  {Segmentation},'' in {\em Brainlesion: {Glioma}, {Multiple} {Sclerosis},
  {Stroke} and {Traumatic} {Brain} {Injuries}} (A.~Crimi, B.~Menze, O.~Maier,
  M.~Reyes, S.~Winzeck, and H.~Handels, eds.), (Cham), pp.~138--149, Springer
  International Publishing, 2016.

\bibitem{kamnitsas_efficient_2017}
K.~Kamnitsas, C.~Ledig, V.~F.~J. Newcombe, J.~P. Simpson, A.~D. Kane, D.~K.
  Menon, D.~Rueckert, and B.~Glocker, ``Efficient multi-scale {3D} {CNN} with
  fully connected {CRF} for accurate brain lesion segmentation,'' {\em Medical
  Image Analysis}, vol.~36, pp.~61--78, Feb. 2017.

\bibitem{chen_voxresnet:_2018}
H.~Chen, Q.~Dou, L.~Yu, J.~Qin, and P.-A. Heng, ``{VoxResNet}: {Deep} voxelwise
  residual networks for brain segmentation from {3D} {MR} images,'' {\em
  NeuroImage}, vol.~170, pp.~446--455, Apr. 2018.

\bibitem{chang_brain_2018}
J.~Chang, X.~Zhang, J.~Chang, M.~Ye, D.~Huang, P.~Wang, and C.~Yao, ``Brain
  {Tumor} {Segmentation} {Based} on {3D} {Unet} with {Multi}-{Class} {Focal}
  {Loss},'' in {\em 2018 11th {International} {Congress} on {Image} and
  {Signal} {Processing}, {BioMedical} {Engineering} and {Informatics}
  ({CISP}-{BMEI})}, pp.~1--5, Oct. 2018.

\bibitem{maier_isles_2017}
O.~Maier, B.~H. Menze, J.~von~der Gablentz, L.~Häni, M.~P. Heinrich,
  M.~Liebrand, S.~Winzeck, A.~Basit, P.~Bentley, L.~Chen, D.~Christiaens,
  F.~Dutil, K.~Egger, C.~Feng, B.~Glocker, M.~Götz, T.~Haeck, H.-L. Halme,
  M.~Havaei, K.~M. Iftekharuddin, P.-M. Jodoin, K.~Kamnitsas, E.~Kellner,
  A.~Korvenoja, H.~Larochelle, C.~Ledig, J.-H. Lee, F.~Maes, Q.~Mahmood, K.~H.
  Maier-Hein, R.~McKinley, J.~Muschelli, C.~Pal, L.~Pei, J.~R. Rangarajan,
  S.~M.~S. Reza, D.~Robben, D.~Rueckert, E.~Salli, P.~Suetens, C.-W. Wang,
  M.~Wilms, J.~S. Kirschke, U.~M. Krämer, T.~F. Münte, P.~Schramm, R.~Wiest,
  H.~Handels, and M.~Reyes, ``{ISLES} 2015 - {A} public evaluation benchmark
  for ischemic stroke lesion segmentation from multispectral {MRI},'' {\em
  Medical Image Analysis}, vol.~35, pp.~250--269, Jan. 2017.

\bibitem{menze_multimodal_2015}
B.~H. Menze, A.~Jakab, S.~Bauer, J.~Kalpathy-Cramer, K.~Farahani, J.~Kirby,
  Y.~Burren, N.~Porz, J.~Slotboom, R.~Wiest, L.~Lanczi, E.~Gerstner, M.~Weber,
  T.~Arbel, B.~B. Avants, N.~Ayache, P.~Buendia, D.~L. Collins, N.~Cordier,
  J.~J. Corso, A.~Criminisi, T.~Das, H.~Delingette, Ã.~Demiralp, C.~R. Durst,
  M.~Dojat, S.~Doyle, J.~Festa, F.~Forbes, E.~Geremia, B.~Glocker, P.~Golland,
  X.~Guo, A.~Hamamci, K.~M. Iftekharuddin, R.~Jena, N.~M. John, E.~Konukoglu,
  D.~Lashkari, J.~A. Mariz, R.~Meier, S.~Pereira, D.~Precup, S.~J. Price, T.~R.
  Raviv, S.~M.~S. Reza, M.~Ryan, D.~Sarikaya, L.~Schwartz, H.~Shin, J.~Shotton,
  C.~A. Silva, N.~Sousa, N.~K. Subbanna, G.~Szekely, T.~J. Taylor, O.~M.
  Thomas, N.~J. Tustison, G.~Unal, F.~Vasseur, M.~Wintermark, D.~H. Ye,
  L.~Zhao, B.~Zhao, D.~Zikic, M.~Prastawa, M.~Reyes, and K.~V. Leemput, ``The
  {Multimodal} {Brain} {Tumor} {Image} {Segmentation} {Benchmark} ({BRATS}),''
  {\em IEEE Transactions on Medical Imaging}, vol.~34, pp.~1993--2024, Oct.
  2015.

\bibitem{Ronneberger2015}
O.~Ronneberger, P.~Fischer, and T.~Brox, ``U-net: Convolutional networks for
  biomedical image segmentation,'' 2015.

\bibitem{Chen2018}
Y.~Chen, S.~Liu, Y.~Kang, and E.~M. Haacke, ``{A rapid, robust multi-echo phase
  unwrapping method for quantitative susceptibility mapping (QSM) using
  strategically acquired gradient echo (STAGE) data acquisition},'' in {\em
  Medical Imaging 2018: Physics of Medical Imaging} (J.~Y. Lo, T.~G. Schmidt,
  and G.-H. Chen, eds.), vol.~10573, pp.~735 -- 742, International Society for
  Optics and Photonics, SPIE, 2018.

\bibitem{Paszke2017}
A.~Paszke, S.~Gross, S.~Chintala, G.~Chanan, E.~Yang, Z.~DeVito, Z.~Lin,
  A.~Desmaison, L.~Antiga, and A.~Lerer, ``Automatic differentiation in
  pytorch,'' 2017.

\bibitem{Lowekamp2013}
B.~Lowekamp, D.~Chen, L.~Ibanez, and D.~Blezek, ``The design of simpleitk,''
  {\em Frontiers in Neuroinformatics}, vol.~7, p.~45, 2013.

\bibitem{Yushkevich2006}
P.~A. Yushkevich, J.~Piven, H.~Cody~Hazlett, R.~Gimpel~Smith, S.~Ho, J.~C. Gee,
  and G.~Gerig, ``User-guided {3D} active contour segmentation of anatomical
  structures: Significantly improved efficiency and reliability,'' {\em
  Neuroimage}, vol.~31, no.~3, pp.~1116--1128, 2006.

\bibitem{Kingma2014}
D.~P. Kingma and J.~Ba, ``Adam: A method for stochastic optimization,'' 2014.

\end{thebibliography}

\end{document}